\documentclass{article} 
\usepackage{iclr2020_conference,times}

\usepackage{amsmath,amsfonts,bm}

\def\eqref#1{equation~\ref{#1}}

\def\1{\bm{1}}

\DeclareMathAlphabet{\mathsfit}{\encodingdefault}{\sfdefault}{m}{sl}
\SetMathAlphabet{\mathsfit}{bold}{\encodingdefault}{\sfdefault}{bx}{n}

\usepackage{graphicx}
\usepackage{multirow}
\usepackage{subcaption}
\usepackage{hyperref}
\usepackage{url}
\hypersetup{
    colorlinks=true,
    linkcolor=blue,
    filecolor=magenta,      
    urlcolor=cyan,
}

\usepackage[normalem]{ulem}

\newcommand{\Dataset}{ \mathcal{D} }
\newcommand{\reals}{ \mathbb{R} }

\title{Parameters Estimation from the 21 cm signal using Variational Inference}

\author{H\'ector J. Hort\'ua, Riccardo Volpi \& Luigi  Malag\`o\\
Machine Learning and Optimization Group,\\
Romanian Institute of Science and Technology (RIST),\\
Cluj-Napoca, Romania\\
\texttt{\{hortua.orjuela,volpi,malago\}@rist.ro} \\
}

\iclrfinalcopy 
\begin{document}
\maketitle
\begin{abstract}

Upcoming experiments such as Hydrogen Epoch of Reionization Array (HERA) and Square Kilometre Array (SKA) are intended to measure the 21cm signal over a wide range of redshifts, representing an incredible opportunity in advancing our understanding about the nature of cosmic reionization. At the same time these kind of experiments will present new challenges in processing the extensive amount of data generated, calling for the development of automated methods capable of precisely estimating physical parameters and their uncertainties. In this paper we employ Variational Inference, and in particular Bayesian Neural Networks,   as an alternative to MCMC in 21 cm observations to report credible estimations for cosmological and astrophysical parameters  and assess the correlations among them.
\end{abstract}
 
\section{Introduction}
In the past decade, Cosmology has entered into a new precision era also due to the considerable number of experiments performed to obtain information both from early stages of the Universe through the Cosmic Microwave Background (CMB) and late times via deep redshift surveys of large-scale structures. These measurements have  yielded precise  estimates  for  the   parameters  in the standard  cosmological model, establishing  the current understanding of the Universe. However, the intermediate  time known as  Epoch of Reionization (EoR),  when the first stars and galaxies  ionized  the InterGalactic Medium (IGM), remains vastly unexplored. This period is relevant  to understand the properties of the first structures of our Universe and provide complementary information  related to fundamental Cosmology, inflationary models, and neutrino constraints, among others, e.g.,~\cite{Pritchard_2012}. It is known that EoR can be studied indirectly through its imprint in the IGM, using the redshifted 21cm line~\cite{10.1093/mnras/stv571}. 
This line results from  hyperfine splitting of the  ground state due to the coupled magnetic moments between the proton and the electron in a hydrogen atom, emitting radiation with a 21cm wavelength,  and then redshifted  by the expansion of the Universe~\cite{Pritchard_2012}. 
Future experiments such as Hydrogen Epoch of Reionization Array (HERA)\footnote{https://reionization.org/} and the  Square Kilometre Array (SKA)\footnote{https://www.skatelescope.org/} are intended to measure this 21cm signal over a wide range of redshifts  providing 3D maps of the first hundreds millions years of the Universe. These instruments are expected to generate a huge amount of spectra, encouraging the development of automated methods capable of reliably estimating physical parameters with great accuracy.

Recently, Deep Neural Networks (DNNs) have been applied in several fields of Astronomy because of their ability to extract complex information from data and  efficiently  solve  nonlinear  inversion  problems. In particular, the application of DNN to 21cm signal has received considerable attention due to the success of classifying reionization models~\citep{10.1093/mnras/sty3282} or estimating physical parameters~\citep{10.1093/mnras/stz010}. 
For example, in~\cite{10.1093/mnras/stz010} 2D images corresponding to  slices along the line-of-sight axis of the light-cones were used for training convolutional
neural networks (CNN) in order to estimate some astrophysical parameters. More recently~\cite{La_Plante_2019} and~\cite{hassan2019constraining} generalized the previous findings by incorporating  observational foregrounds expected from future experiments.
However, DNNs are prone to  over-fitting due to the high number of parameters to be adjusted, and
do not provide a measure of uncertainty for the estimated parameters, see for instance~\cite{NIPS2011_4329,KWON2020106816,Cobb_2019}.
This problem can be addressed by following a Bayesian approach  that  allows to quantify the uncertainty in the predicted parameters and provide estimates for the ignorance of the model.

In this paper, we generalize previous work related to the application of DNNs on 21cm data by implementing Bayesian Neural Network (BNNs) to obtain reliable estimates.
The main contributions of this paper are the following:
\begin{enumerate}
    \item Present the first study that employs BNNs for the 21cm dataset, motivating the use of  variational inference techniques in this field.
    \item Show that BNNs applied to 21cm data outperform traditional approaches being able to provide estimations for uncertainty and correlations among the predicted physical parameters.
\end{enumerate}

\subsection{Variational Inference}
BNNs provide the adequate groundwork to output reliable estimations for many  machine learning tasks. Let us consider a training dataset $\Dataset=\{(\bm{x}_i, \bm{y}_i)\}_{i=1}^D$ formed by $D$ couples of images $\bm{x}_i \in \reals^M $  and their respective targets $\bm{y}_i \in \reals^N$. Setting a prior distribution $p({\bm w})$ on the model parameters ${\bm w}$, the posterior distribution can be obtained from Bayes law as $ p({\bm w}|\mathcal{D})\sim p(\mathcal{D}|{\bm w})p({\bm w})$.
Unfortunately, the  posterior usually cannot  be obtained analytically and thus approximate methods are commonly used to perform the inference task. The Variational Inference approach approximates the exact posterior  $p({\bm w}|\mathcal{D})$ by a parametric distribution $q({\bm w}|\theta)$  depending on a set of variational parameters ${\bm \theta}$. These parameters are adjusted to minimize a certain loss function, usually given by the KullBack-Leibler divergence  $\text{KL}(q({\bm w}|{\bm \theta})||p({\bm w}|\mathcal{D}))$. It has been shown that minimizing the KL divergence is equivalent to minimizing the following objective  function~\citep{NIPS2011_4329}
\begin{equation}\label{eq:4}
\mathcal{F}_{VI}(\mathcal{D},{\bm \theta}) = \text{KL}(q({\bm w}|{\bm \theta})||p({\bm w}))\\
-\sum_{(\bm{x},\bm{y})\in\Dataset}\int_\Omega q({\bm w}|{\bm \theta}) \ln p({\bm y}|{\bm x},{\bm w}) d{\bm w} \; .
\end{equation}
Let ${\hat{{\bm \theta}}}$ be the value of ${\bm \theta}$ after training, i.e., corresponding to a minimum of $\mathcal{F}_{VI}(\mathcal{D},{\bm \theta})$. The approximate predictive distribution  $q_{\hat{{\bm \theta}}}$ of $\bm{y}^*$ for a new input $\bm{x}^*$ can be rewritten as~\citep{Gal2015BayesianCN}
\begin{equation}\label{eq:approxpost_pygivenx}
  q_{\hat{{\bm \theta}}}({\bm y}^*|{\bm x}^*)=\int_{\Omega} p({\bm y}^*|{\bm x}^*,{\bm w})q({\bm w}|{\hat{\bm\theta}})d{\bm w} \; .
\end{equation}
Moreover~\cite{Gal2015Dropout} proposed an unbiased Monte-Carlo estimator for Eq.~\ref{eq:approxpost_pygivenx}
\begin{equation}\label{eq:5}
 q_{\hat{{\bm \theta}}}({\bm y}^*|{\bm x}^*) \approx \frac{1}{K}\sum_{k=1}^K p({\bm y}^*|{\bm x}^*,\hat{{\bm w}}_k), \quad \mbox{with } \hat{{\bm w}}_k \sim q({\bm w}|{\hat{\bm\theta}}) \;,
\end{equation}
where $K$ is the number of samples. To infer the correlations between the  parameters, as done in~\cite{2019arXiv191108508H}, we need to predict the full covariance matrix. This requires to produce in output of the last layer of the network a mean vector $\bm{\mu}\in\mathbb{R}^{N}$ and a the covariance matrix $\Sigma\in\mathbb{R}^{N(N+1)/2}$ that represents the aleatoric uncertainty. These outputs  will form  the negative log-likelihood(NLL) for a Multivariate Gaussian distributions~\cite{Dorta_2018,Cobb_2019,2019arXiv191108508H}.
In Bayesian deep learning~\citep{NIPS2017_7141} two main uncertainties are of interest: the aleatoric, capturing the inherent noise in the input data, and epistemic one, capturing the uncertainty in the model, typically due to the lack of data points similar to the current observation.
To obtain both uncertainties~\cite{KWON2020106816},
the images are forward passed through the network $T$ times, obtaining a set of mean vectors $\bm{\mu}_t$ and a covariance matrices $\Sigma_t$. Then, an estimator for the total covariance of the trained model
can be written as  
\begin{equation}\label{eq:cov}
\mathrm{Cov}_{q_{\hat{{\bm \theta}}}}({\bm y}^*,{\bm y}^*|{\bm x}^*)
\approx
\underbrace{\frac{1}{T}\sum_{t=1}^{T}\Sigma_t}_\text{Aleatoric}+ \underbrace{\frac{1}{T}\sum_{t=1}^{T}( {{\bm{\mu}}}_{t}-\bm{\overline{\mu}})( {\bm{\mu}}_{t}-\bm{\overline{\mu}})^{\mathrm{T}}}_\text{Epistemic}, 
\end{equation}
with $\bm{\overline{\mu}}= \frac{1}{T}\sum_{t=1}^{T} {\bm{\mu}}_t$. In this setting, DNNs  can be used to learn the correlations between the targets and to produce estimates of their uncertainties. 
\begin{figure}[htb]
\begin{center}
\includegraphics[width=0.7\linewidth]{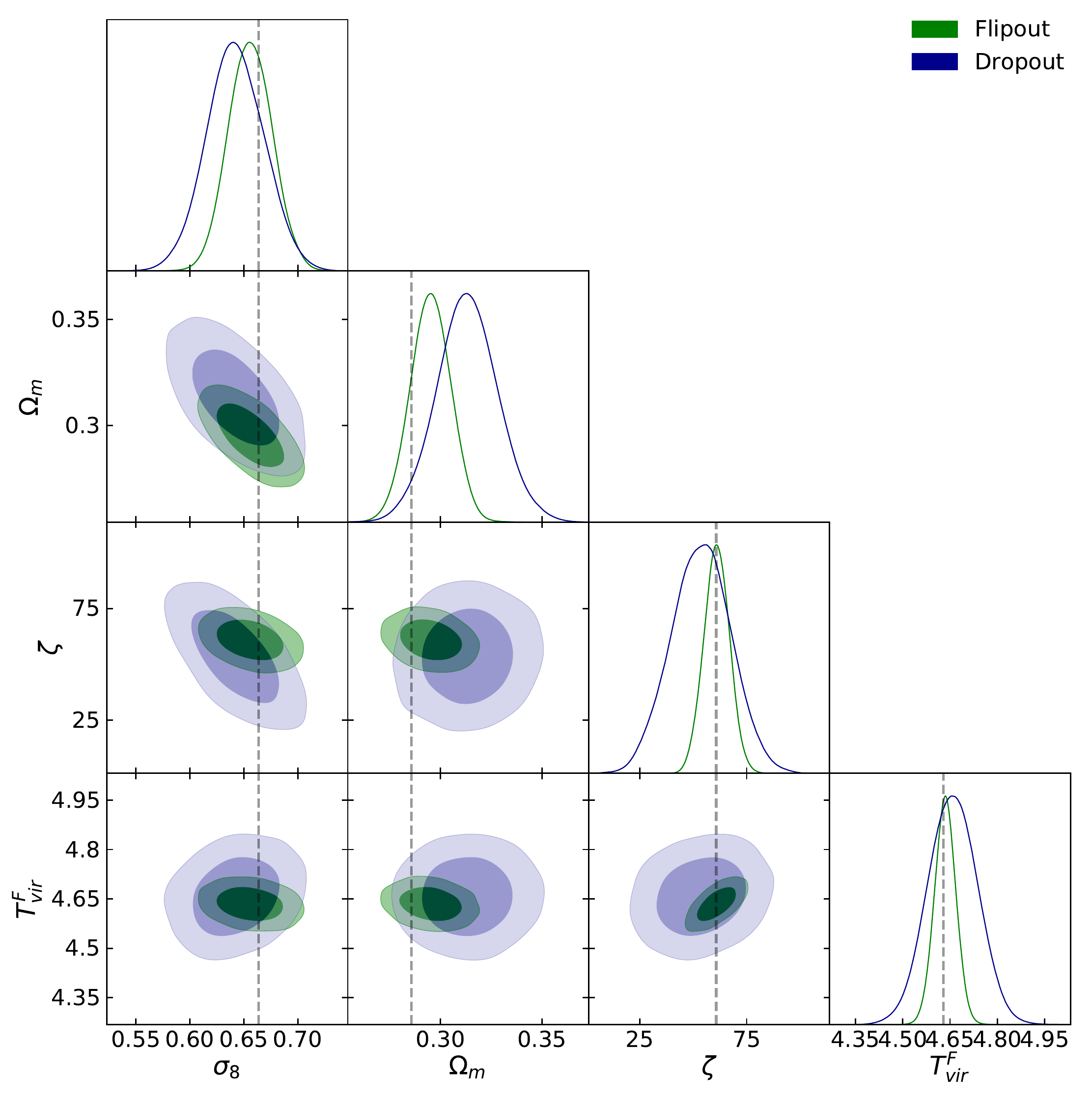}
\end{center}
\caption{Posterior distributions of the parameters for one example in the test set. The dashed lines stand for the real values. The contour regions in the two-dimensional posteriors stand for 68 and 95$\%$ confidence levels.}
\label{fig:bestCIsp}
\end{figure}
\begin{table}[h]
\centering
\begin{tabular}{|l|c|c|c|c|}
\hline 
                  & $\sigma_8$     & $\Omega_m$    &  $\zeta$   & $T^F_{vir}$    \\ \hline
Flipout              &   $0.656^{+0.040}_{-0.039}   $   &  $0.295^{+0.019}_{-0.020}   $    &   $61.00^{+10.00}_{-10.00}            $   & $4.634^{+0.070}_{-0.068}   $     \\ \hline
Dropout              &  $0.643^{+0.052}_{-0.051}$    &   $0.313^{+0.030}_{-0.029}$   &  $53.01^{+30.00}_{-30.00}$    & $4.660^{+0.150}_{-0.160}$      \\ \hline
Example value  & $0.663$ &$0.285$ &$60.74 $&$4.630$\\ \hline
\end{tabular}
\caption{Limits at the $95\%$ confidence level of the credible interval of predicted parameters.}
\label{table:bestCIsp}
\end{table}

\section{Results}
\subsection{Dataset}
We generated 21cm simulations through the semi-numerical code   21cmFast\footnote{https://21cmfast.readthedocs.io/en/latest/}, producing realizations of halo distributions and ionization maps at high redshifts. We varied four parameters in total to produce the dataset. Two parameters corresponding to the cosmological context: the matter density parameter $\Omega_m \in [0.2,0.4]$ and the rms linear fluctuation in the mass distribution on $8h^{-1}$Mpc $\sigma_8 \in [0.6,0.8]$. The other two parameters corresponding to the astrophysical context: the ionizing efficiency of high-z galaxies $\zeta \in [10,100]$ and the minimum virial temperature of star-forming haloes $T^F_{vir} \in [3.98,39.80]\times10^4$K (hereafter represented in log10 units). For each set of parameters we have produced 20 images at different redshifts in the range $z \in [6,12]$, and stacked these redshift-images into a single multi-channel tensor.  This scheme brings two main advantages, first the network can extract effectively the information encoded over images as it was reported in \cite{2019arXiv191108508H}, and secondly, it represents adequately the signals for the  next-generation interferometers and provides advantage when we need to include effects of foreground contamination  \cite{La_Plante_2019}.
As a final result we have obtained $6,000$ images each with size of $(128,128,20)$ and resolution $1.5$ Mpc. We used a 70-10-20 split for training, validation and test, respectively.

\subsection{Architecture and Evaluation}\label{gen_inst}
We used a modified version of the VGG architecture with 5 VGG blocks (each made by two Conv2D layers and one max pooling) and channels size [32, 32, 32, 32, 64]. Kernel size is fixed to 3$\times$3 and activation function used is LeakyReLU ($\alpha=-0.3$). Each convolutional layer in the network is followed by a batch renormalization layer. 
We trained the networks for 200 epochs with batches of 32 samples. To obtain the approximated posterior and the related uncertainties, we feed each input image from the test set 3,500 times to each network.
To model the distribution over the weights we used two methods, Dropout~\citep{Gal2015Dropout} and Flipout~\citep{wen2018flipout}. For Dropout we tested several dropout rates in the range [$0.01$, $0.1$] while keeping L2 regularization fixed to $1e^{-5}$, while for Flipout we tested several L2 regularizations in the range [$1e^{-5}$, $1e^{-7}$]. We found the following configurations to be the best performing combinations of hyperparameters: Flipout with L2 regularizer $1e^{-7}$ and Dropout with dropout rate $1e^{-2}$. Hereafter we will report the best hyperparameters combination, remaining experiments are reported in Fig.~\ref{fig:all4CI} and Tables~\ref{table:Dropoutall}-\ref{table:Flipoutall} in Appendix.
In Fig.~\ref{fig:bestCIsp} we report the confidence intervals\footnote{We use the getdist~\cite{Lewis:2019xzd} package to produce the plots.} for a single example in the test set and in Table~\ref{table:bestCIsp} we present the parameter $68\%$ confidence level. Notice that Flipout  yields more accurate inferences and provides tighter constraints contours, see for example $T^F_{vir}$-$\zeta$. Moreover, the correlations extracted from EoR such as $\sigma_8$-$\Omega_m$ (see Fig.~\ref{fig:bestCIsp}) provide significant information for breaking  parameter degeneracies   and thus,  be able  to improve the existing  measurements on cosmological parameters~\cite{Pritchard_2012,McQuinn_2006}.
To evaluate the performance of the network in terms of predicting the parameters,  we compute the coefficient of determination 
\begin{equation}
R^2=1-\frac{\sum_i (\bm{\bar{\mu}}(\bm{x}_i)-\bm{y}_i)^2}{\sum_i (\bm{y}_i-\bm{\bar{y}})^2}
\end{equation}
where $\bm{\bar{\mu}}(\bm{x}_i)$ (see Eq.~\ref{eq:cov}) is the prediction of the trained Bayesian network , $\bm{\bar{y}}$ is the average of the true parameters and the summations are performed over the entire test set. $R^2$ ranges from 0 to 1, where 1 represents perfect inference.
Table~\ref{table:bestCI} reports the coefficient of determination and average confidence intervals for Flipout and Dropout after calibration~\cite{2019arXiv191108508H}. Flipout obtains the best estimations even though tends to overestimate its uncertainties. The results obtained are comparable with the state of the art~\cite{10.1093/mnras/stz010,hassan2019constraining} even if the architecture used in the present paper is considerably smaller, furthermore as discussed in this section our approach has the advantage to be able to provide reliable uncertainty estimation for the predicted parameters. 
\begin{table}[h]
\begin{center}
\begin{tabular}{|c|c|c|c|c|c|c|c|c|}
\hline
    &  \multicolumn{4}{c|}{Flipout (NLL=-2.4)}    & \multicolumn{4}{c|}{Dropout (NLL=-0.74)} \\\cline{2-9}
 & $\sigma_8$    &  $\Omega_m$   &  $\zeta$    & $T^F_{vir}$    &   $\sigma_8$   &   $\Omega_m$   & $\zeta$    & $T^F_{vir}$    \\ \hline
$R^2$                & 0.92 &0.97     &0.83      &0.97     &    0.87  &0.94      &0.65     &0.92    \\ \hline
C.L. $68.3\%$                &   74.1 &70.2  &75.4     &70.3      & 70.4     &  67.3    &58.5  &76.1  \\ \hline
C.L. $95.5\%$                &   97.4   &96.2      &98.5      &96.0      &  95.7    &    96.3  & 91.7     & 98.5     \\ \hline
C.L. $99.7\%$                &  99.7    &99.9      &99.9      &99.9      & 99.6     &    99.8  & 99.8     &99.9      \\ \hline
\end{tabular}
\caption{Metrics for the best experiments with Flipout and Dropout.}
\label{table:bestCI}
\end{center}
\end{table}
\section{Conclusions}
We presented the first study using a Bayesian approach to obtain credible estimates for astrophysical and cosmological parameters from 21cm signals. Flipout outperforms Dropout and is able both to better estimate correlations and to obtain a better coefficient of determination.
We obtain performances comparable with the existing literature, while using a relatively smaller network~\cite{10.1093/mnras/stz010,hassan2019constraining}. By using a BNN based on Variational Inference, our method allows us to estimate confidence intervals for the predictions and parameters correlations. We plan on evaluating the performances of different  network architectures (also in particular residual networks) and estimate the cosmological and astrophysical parameters in the presence of realistic noise from instruments of the future 21 cm surveys. 

\subsubsection*{Acknowledgments}
H.J.~Hort\'ua, R.~Volpi, and L.~Malag\`o are supported by the DeepRiemann project, co-funded by the European Regional Development Fund and the Romanian Government through the Competitiveness Operational Programme 2014-2020, Action 1.1.4, project ID P\_37\_714, contract no. 136/27.09.2016. 

\bibliography{iclr2020_conference}

\begin{thebibliography}{18}
\providecommand{\natexlab}[1]{#1}
\providecommand{\url}[1]{\texttt{#1}}
\expandafter\ifx\csname urlstyle\endcsname\relax
  \providecommand{\doi}[1]{doi: #1}\else
  \providecommand{\doi}{doi: \begingroup \urlstyle{rm}\Url}\fi

\bibitem[Cobb et~al.(2019)Cobb, Himes, Soboczenski, Zorzan, O'Beirne, Baydin,
  Gal, Domagal-Goldman, Arney, and and]{Cobb_2019}
Adam~D. Cobb, Michael~D. Himes, Frank Soboczenski, Simone Zorzan, Molly~D.
  O'Beirne, At{\i}l{\i}m~Güne{\c{s}} Baydin, Yarin Gal, Shawn~D.
  Domagal-Goldman, Giada~N. Arney, and Daniel~Angerhausen and.
\newblock An ensemble of bayesian neural networks for exoplanetary atmospheric
  retrieval.
\newblock \emph{The Astronomical Journal}, 158\penalty0 (1):\penalty0 33, 2019.

\bibitem[Dorta et~al.(2018)Dorta, Vicente, Agapito, Campbell, and
  Simpson]{Dorta_2018}
Garoe Dorta, Sara Vicente, Lourdes Agapito, Neill D.~F. Campbell, and Ivor
  Simpson.
\newblock Structured uncertainty prediction networks.
\newblock \emph{2018 IEEE/CVF Conference on Computer Vision and Pattern
  Recognition}, 2018.

\bibitem[Gal \& Ghahramani(2015{\natexlab{a}})Gal and
  Ghahramani]{Gal2015BayesianCN}
Yarin Gal and Zoubin Ghahramani.
\newblock Bayesian convolutional neural networks with bernoulli approximate
  variational inference, 2015{\natexlab{a}}.

\bibitem[Gal \& Ghahramani(2015{\natexlab{b}})Gal and
  Ghahramani]{Gal2015Dropout}
Yarin Gal and Zoubin Ghahramani.
\newblock Dropout as a {B}ayesian approximation: Insights and applications.
\newblock In \emph{Deep Learning Workshop, ICML}, 2015{\natexlab{b}}.

\bibitem[Gillet et~al.(2019)Gillet, Mesinger, Greig, Liu, and
  Ucci]{10.1093/mnras/stz010}
Nicolas Gillet, Andrei Mesinger, Bradley Greig, Adrian Liu, and Graziano Ucci.
\newblock {Deep learning from 21-cm tomography of the cosmic dawn and
  reionization}.
\newblock \emph{Monthly Notices of the Royal Astronomical Society},
  484\penalty0 (1):\penalty0 282--293, 2019.

\bibitem[Graves(2011)]{NIPS2011_4329}
Alex Graves.
\newblock Practical variational inference for neural networks.
\newblock In J.~Shawe-Taylor, R.~S. Zemel, P.~L. Bartlett, F.~Pereira, and
  K.~Q. Weinberger (eds.), \emph{Advances in Neural Information Processing
  Systems 24}, pp.\  2348--2356. Curran Associates, Inc., 2011.

\bibitem[Greig \& Mesinger(2015)Greig and Mesinger]{10.1093/mnras/stv571}
Bradley Greig and Andrei Mesinger.
\newblock {21CMMC: an MCMC analysis tool enabling astrophysical parameter
  studies of the cosmic 21 cm signal}.
\newblock \emph{Monthly Notices of the Royal Astronomical Society},
  449\penalty0 (4):\penalty0 4246--4263, 2015.

\bibitem[Hassan et~al.(2018)Hassan, Liu, Kohn, and
  La Plante]{10.1093/mnras/sty3282}
Sultan Hassan, Adrian Liu, Saul Kohn, and Paul La Plante.
\newblock {Identifying reionization sources from 21cm maps using Convolutional
  Neural Networks}.
\newblock \emph{Monthly Notices of the Royal Astronomical Society},
  483\penalty0 (2):\penalty0 2524--2537, 2018.

\bibitem[Hassan et~al.(2019)Hassan, Andrianomena, and
  Doughty]{hassan2019constraining}
Sultan Hassan, Sambatra Andrianomena, and Caitlin Doughty.
\newblock Constraining the astrophysics and cosmology from 21cm tomography
  using deep learning with the ska, 2019.

\bibitem[{Hortua} et~al.(2019){Hortua}, {Volpi}, {Marinelli}, and
  {Malag{\`o}}]{2019arXiv191108508H}
Hector~J. {Hortua}, Riccardo {Volpi}, Dimitri {Marinelli}, and Luigi
  {Malag{\`o}}.
\newblock {Parameters Estimation for the Cosmic Microwave Background with
  Bayesian Neural Networks}.
\newblock \emph{arXiv e-prints}, art. arXiv:1911.08508, 2019.

\bibitem[Kendall \& Gal(2017)Kendall and Gal]{NIPS2017_7141}
Alex Kendall and Yarin Gal.
\newblock What uncertainties do we need in bayesian deep learning for computer
  vision?
\newblock In I.~Guyon, U.~V. Luxburg, S.~Bengio, H.~Wallach, R.~Fergus,
  S.~Vishwanathan, and R.~Garnett (eds.), \emph{Advances in Neural Information
  Processing Systems 30}, pp.\  5574--5584. Curran Associates, Inc., 2017.

\bibitem[Kwon et~al.(2020)Kwon, Won, Kim, and Paik]{KWON2020106816}
Yongchan Kwon, Joong-Ho Won, Beom~Joon Kim, and Myunghee~Cho Paik.
\newblock Uncertainty quantification using bayesian neural networks in
  classification: Application to biomedical image segmentation.
\newblock \emph{Computational Statistics and Data Analysis}, 142:\penalty0
  106816, 2020.

\bibitem[Levasseur et~al.(2017)Levasseur, Hezaveh, and
  Wechsler]{Perreault_Levasseur_2017}
Laurence~Perreault Levasseur, Yashar~D. Hezaveh, and Risa~H. Wechsler.
\newblock Uncertainties in parameters estimated with neural networks:
  Application to strong gravitational lensing.
\newblock \emph{The Astrophysical Journal}, 850\penalty0 (1):\penalty0 L7, nov
  2017.
\newblock \doi{10.3847/2041-8213/aa9704}.
\newblock URL \url{https://doi.org/10.3847%2F2041-8213%2Faa9704}.

\bibitem[Lewis(2019)]{Lewis:2019xzd}
Antony Lewis.
\newblock {GetDist: a Python package for analysing Monte Carlo samples}.
\newblock 2019.

\bibitem[McQuinn et~al.(2006)McQuinn, Zahn, Zaldarriaga, Hernquist, and
  Furlanetto]{McQuinn_2006}
Matthew McQuinn, Oliver Zahn, Matias Zaldarriaga, Lars Hernquist, and Steven~R.
  Furlanetto.
\newblock Cosmological parameter estimation using 21 cm radiation from the
  epoch of reionization.
\newblock \emph{The Astrophysical Journal}, 653\penalty0 (2):\penalty0
  815--834, dec 2006.
\newblock \doi{10.1086/505167}.
\newblock URL \url{https://doi.org/10.1086%2F505167}.

\bibitem[Plante \& Ntampaka(2019)Plante and Ntampaka]{La_Plante_2019}
Paul~La Plante and Michelle Ntampaka.
\newblock Machine learning applied to the reionization history of the universe
  in the 21 cm signal.
\newblock \emph{The Astrophysical Journal}, 880\penalty0 (2):\penalty0 110,
  2019.

\bibitem[Pritchard \& Loeb(2012)Pritchard and Loeb]{Pritchard_2012}
Jonathan~R Pritchard and Abraham Loeb.
\newblock 21 cm cosmology in the 21st century.
\newblock \emph{Reports on Progress in Physics}, 75\penalty0 (8):\penalty0
  086901, 2012.

\bibitem[Wen et~al.(2018)Wen, Vicol, Ba, Tran, and Grosse]{wen2018flipout}
Yeming Wen, Paul Vicol, Jimmy Ba, Dustin Tran, and Roger Grosse.
\newblock Flipout: Efficient pseudo-independent weight perturbations on
  mini-batches.
\newblock In \emph{International Conference on Learning Representations}, 2018.

\end{thebibliography}
\bibliographystyle{iclr2020_conference}

\appendix
\section{Appendix}
In Tables~\ref{table:Dropoutall}-\ref{table:Flipoutall} we report different experiments, to determine the most adequate technique for estimating the parameters from the 21cm dataset. First, we found that sampling more than once during training improves the results. Second, Flipout does a good job for extracting the information in the 21cm images rather than other techniques such as Dropout.
 \begin{table}[h]
\centering
\begin{tabular}{|l|l|l|l|l|}
\hline
 & \multicolumn{2}{l|}{Dropout dr$=1e^{-2}$} & \multicolumn{2}{l|}{Dropout dr$=0.1$} \\ \cline{2-5} 
                  & Sample=1          & Sample=10          & Sample=1          & Sample=10          \\ \hline
NLL               &    -0.18        & -0.74            &  0.99          & 0.28            \\ \hline
$R^2$                &     0.77       & 0.85            &      0.70      &    0.78         \\ \hline
$68\%$ C.L.         &    65.3        &    68.1         &       66.7     &     65.0        \\ \hline
$95\%$ C.L.     &    94.1        &       95.5      &     92.8       &     92.2        \\ \hline
$99\%$ C.L.     &       99.3     &      99.7       &       98.7     &     98.7        \\ \hline
\end{tabular}
\caption{Metrics for all Dropout experiments: dr$=(1e^{-2},0.1)$, reg $=1e^{-5}$. In each experiment, we sample once and ten times during training.}
\label{table:Dropoutall}
\end{table}
\begin{table}[h]
\centering
\begin{tabular}{|l|l|l|l|l|}
\hline
 & \multicolumn{2}{l|}{Flipout reg$=1e{-7}$} & \multicolumn{2}{l|}{Flipout reg$=1e^{-5}$} \\ \cline{2-5} 
 & Sample=1           & Sample=10          & Sample=1           & Sample=10 \\ \hline
NLL               &    -2.30         & -2.40            &     -1.81        &-2.00   \\ \hline
$R^2$                &    0.91         &   0.92          &     0.84        & 0.84   \\ \hline
$68\%$ C.L.                &     75.5        &  72.5           &     76.2        &  76.4    \\ \hline
$95\%$ C.L.                &       97.2      &  97.0           &     97.6        &      97.5    \\ \hline
$99\%$ C.L.                &      99.5       &        99.8     &    99.8         &      99.8    \\ \hline
\end{tabular}
\caption{Metrics for all Flipout experiments: reg $=(1e^{-5}, 1e^{-7})$. In each experiment, we sample once and ten times during training.}
\label{table:Flipoutall}
\end{table}

\begin{figure}[h]
    \centering
    \begin{subfigure}[b]{0.45\textwidth}
        \includegraphics[width=\textwidth]{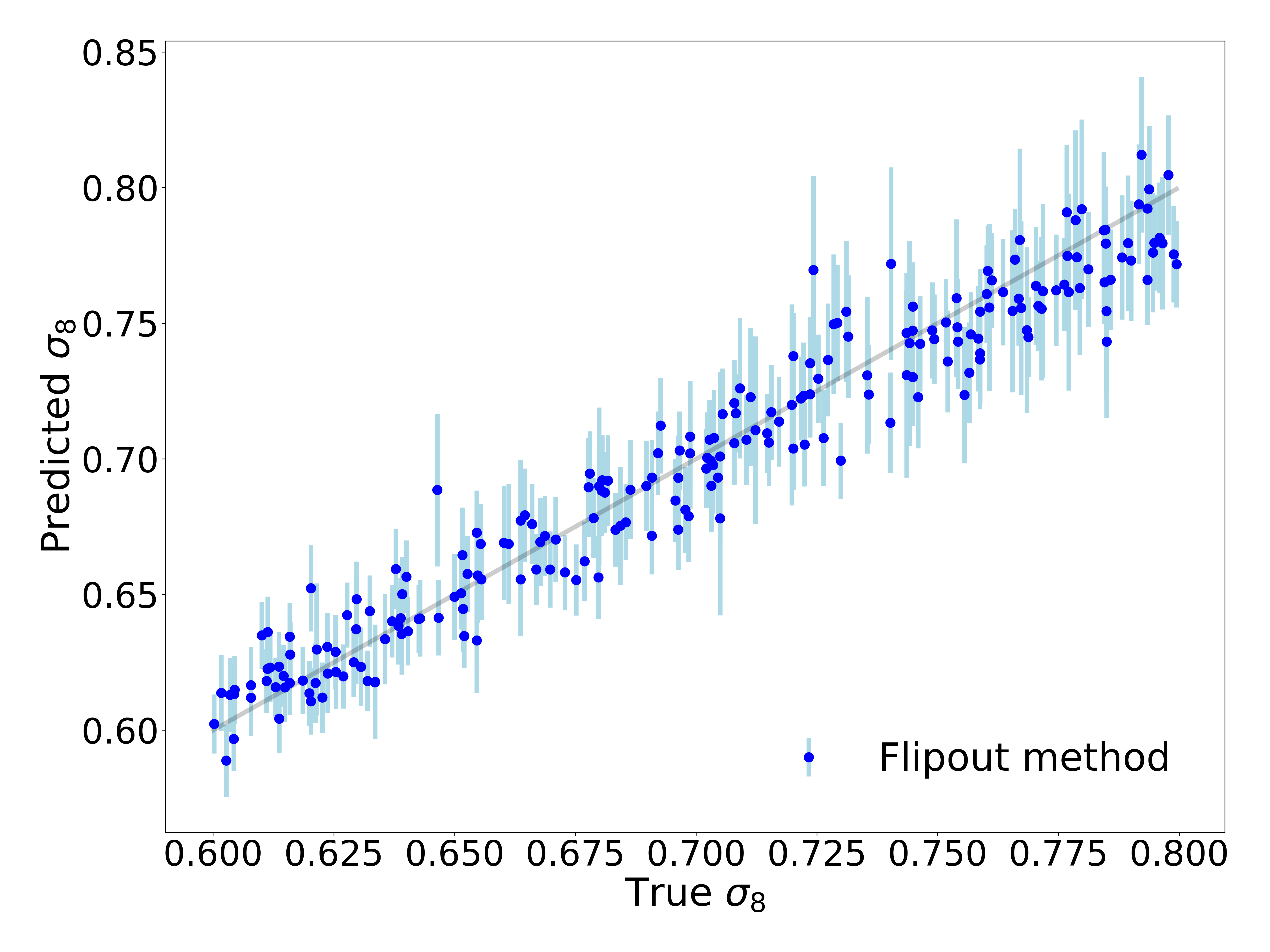}
        \caption{Predicted $\sigma_8$ value vs input (true) value.}
        \label{figpreda}
    \end{subfigure}
    ~ 
    \begin{subfigure}[b]{0.45\textwidth}
        \includegraphics[width=\textwidth]{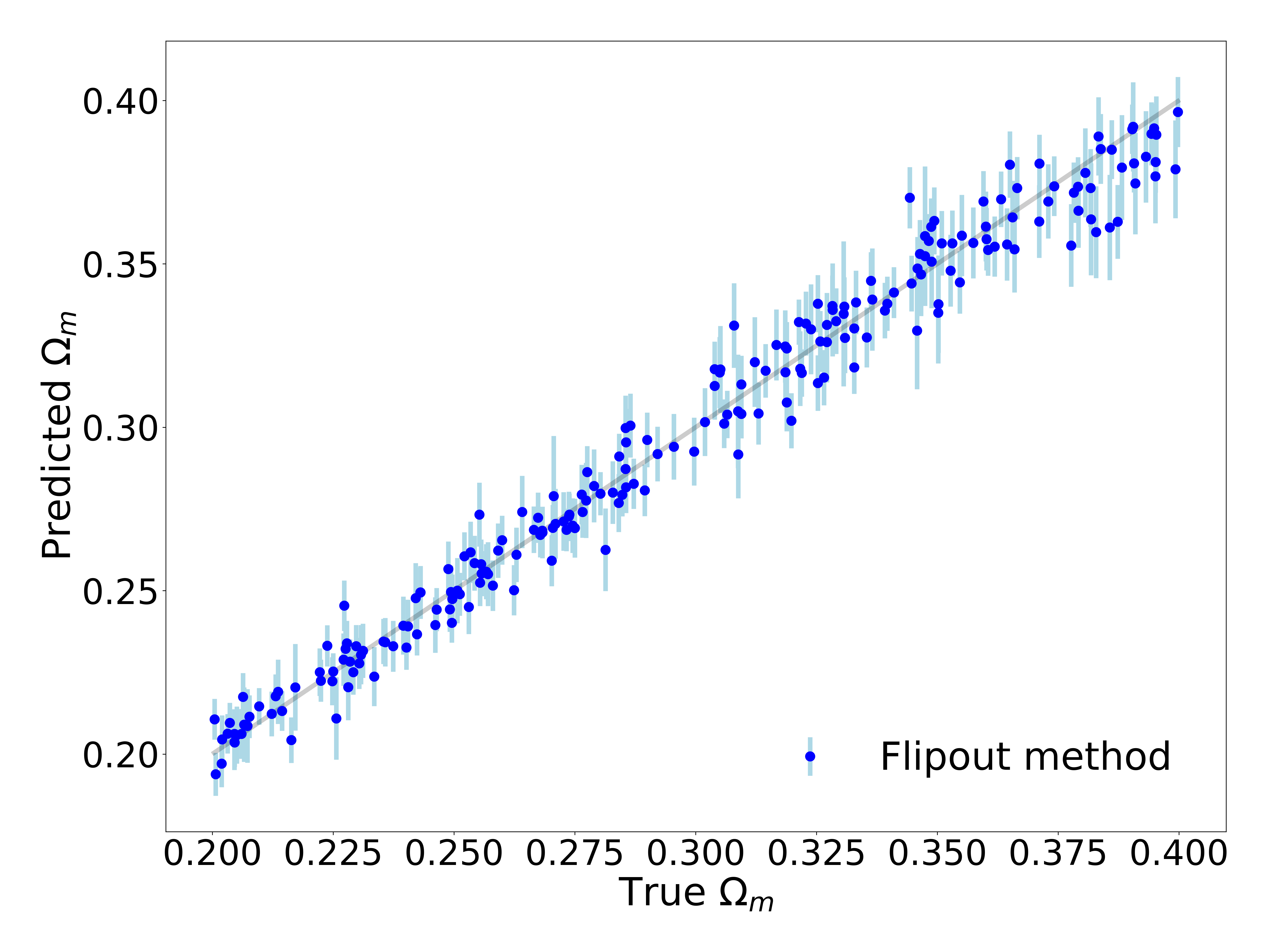}
        \caption{Predicted $\Omega_m$ value vs input (true) value.}
        \label{figpredb}
    \end{subfigure}
    ~ 
    \begin{subfigure}[b]{0.45\textwidth}
        \includegraphics[width=\textwidth]{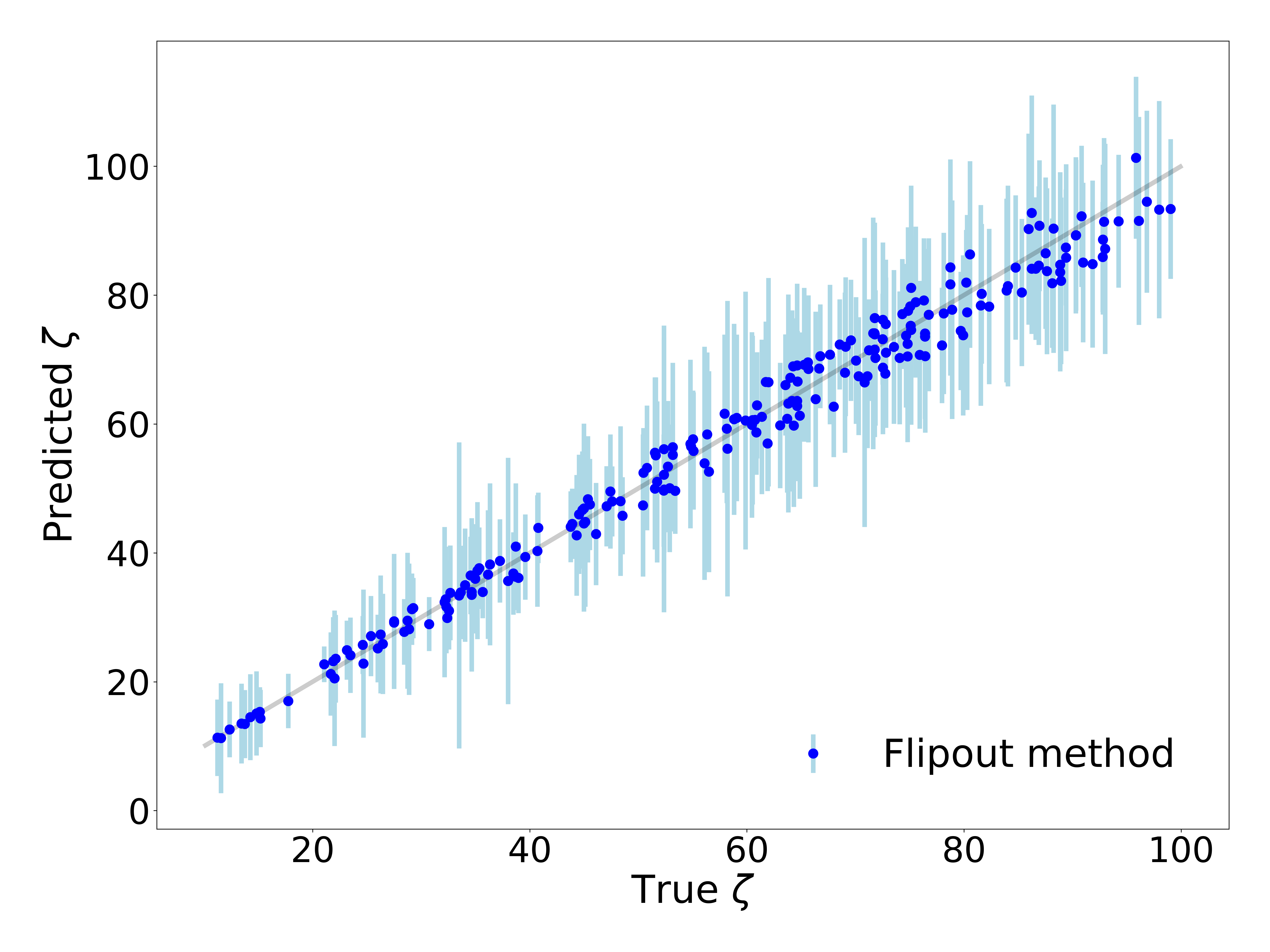}
        \caption{Predicted $\zeta$ value vs input (true) value.}
        \label{figpredc}
    \end{subfigure}
     \begin{subfigure}[b]{0.45\textwidth}
        \includegraphics[width=\textwidth]{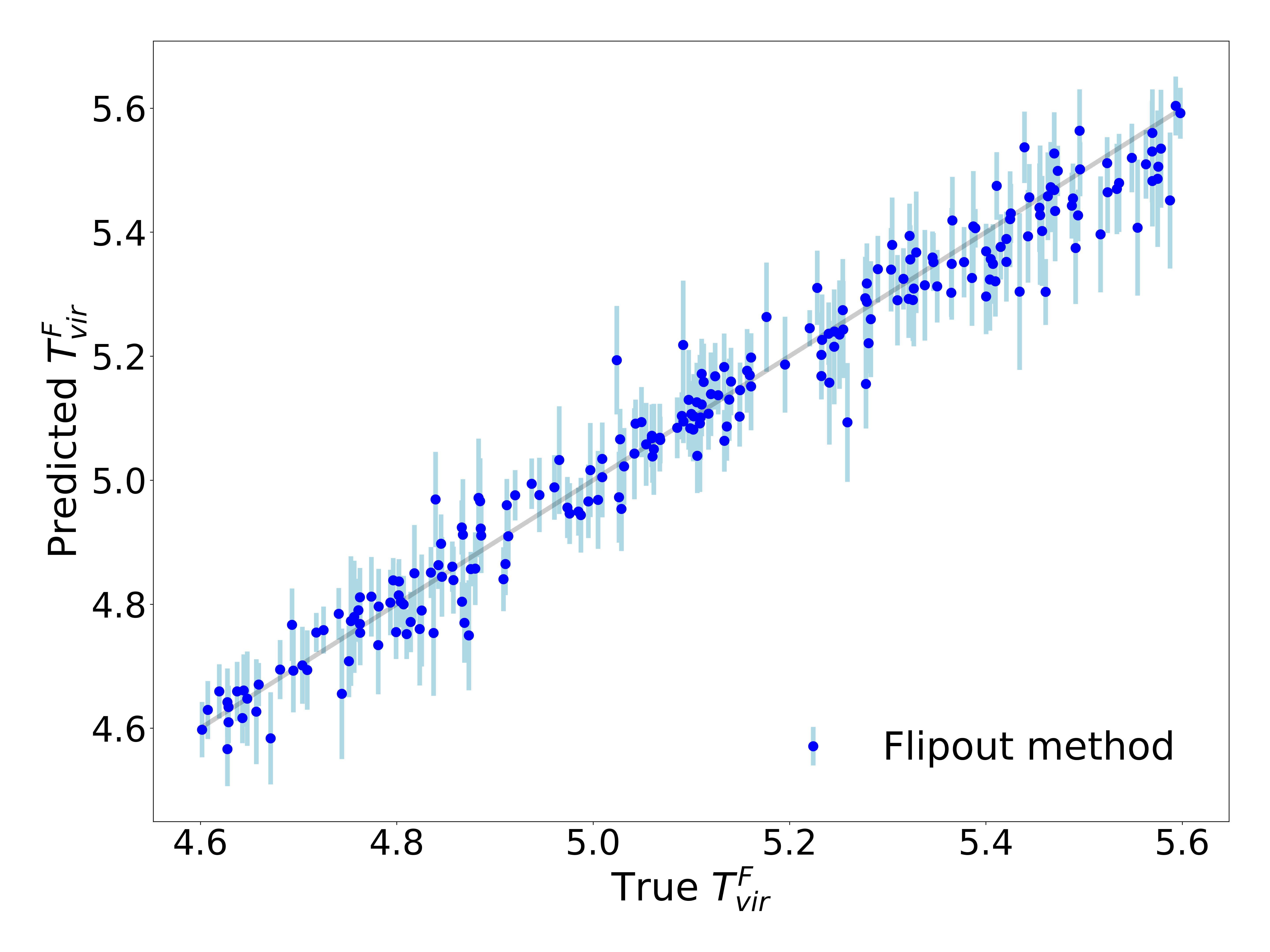}
        \caption{Predicted $T^F_{vir}$ value vs input (true) value.}
        \label{figpredd}
    \end{subfigure}
    \caption{Parameter inference using the test set. The images in each panel contain the true values vs thhe predicted parameter values. The shadow lines represent the total uncertainty, epistemic plus aleatoric.}\label{figapendix1}
\end{figure}

\begin{figure}[h]
\begin{center}
\includegraphics[width=0.8\linewidth]{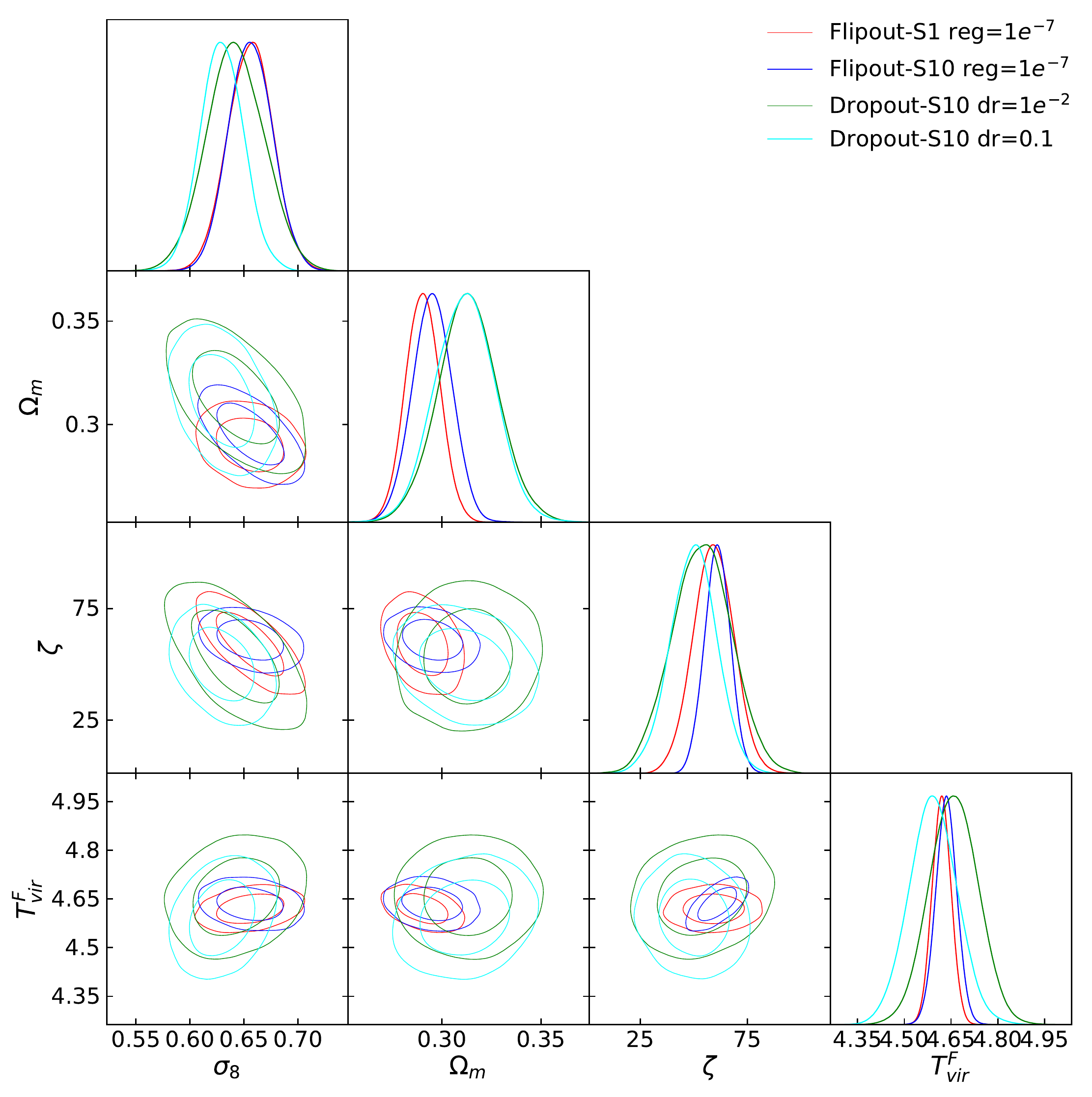}
\end{center}
\caption{One and two-dimensional posterior distributions of the parameters for one example in the test set. Each color represents a different method.}
\label{fig:all4CI}
\end{figure}

Finally, we observe that Dropout underestimates its uncertainties while Flipout overestimates its uncertainties, therefore methods for calibration should be used before reporting the predictions.  To compute the confidence level reported in Tables~\ref{table:Dropoutall}-\ref{table:Flipoutall},  we have binned the 3,500 samples and computed the mode \cite{Perreault_Levasseur_2017}.  With this value, and  assuming an  unimodal posterior, we estimated the intervals that include the 68, 95, and 99$\%$ of the samples. The contour regions for the best results are also reported in Fig.~\ref{fig:all4CI}. In Fig.~\ref{figapendix1} we plot the predicted and true values of the cosmological and astrophysical parameters using Flipout. The $R^2$ for this method is reported  in Table~\ref{table:Flipoutall}. 

\end{document}